\documentclass[aps,prl,10pt,twocolumn]{revtex4-1}
\usepackage{graphicx}
\usepackage{amsmath}
\usepackage{amssymb}
\usepackage{mathrsfs}
\usepackage{color}
\usepackage{ulem}
\usepackage{multirow}
\usepackage{epstopdf}

\begin{document}

	\title{``Molecular waveplate" for the control of ultrashort pulses carrying orbital angular momentum}
	\author{Chengqing Xu$^1$}
\author{Lixin He$^{1}$}
\email{helx\_hust@hust.edu.cn}

\author{Wanchen Tao$^1$, Xiaosong Zhu$^1$}

\author{Feng Wang$^2$}

\author{Long Xu$^3$}

\author{Lu Xu$^1$}
\email{luxu\_0909@hust.edu.cn}

\author{Pengfei Lan$^{1}$}
\email{pengfeilan@hust.edu.cn}

\author{Ilya Averbukh$^{4,5}$}

\author{Yehiam Prior$^4$}

\author{Peixiang Lu$^{1,2,}$}
\email{lupeixiang@hust.edu.cn}

\affiliation{%
	$^1$Wuhan National Laboratory for Optoelectronics and School of Physics, Huazhong University of Science and Technology, Wuhan 430074, China\\
	$^2$Hubei Key Laboratory of Optical Information and Pattern Recognition, Wuhan Institute of Technology, Wuhan 430205, China.\\
	$^3$Department of Physics, Xiamen University, Xiamen 361005, China.\\
	$^4$AMOS and Department of Chemical and Biological Physics, The Weizmann Institute of Science, Rehovot 7610001, Israel.\\
	$^5$Department of Chemistry, University of British Columbia, Vancouver, British Columbia V6T 1Z1, Canada.\\
}%

	
\begin{abstract}
Ultrashort laser pulses carrying orbital angular momentum (OAM) have become essential tools in Atomic, Molecular, and Optical (AMO) studies, particularly for investigating strong-field light-matter interactions.  However, controlling and generating ultrashort vortex pulses presents significant challenges, since their broad spectral content complicates manipulation with conventional optical elements, while the high peak power inherent in short-duration pulses risks damaging optical components.
Here, we introduce a novel method for generating and controlling broadband ultrashort vortex beams by exploiting the non-adiabatic alignment of linear gas-phase molecules induced by vector beams. The interaction between the vector beam and the gas-phase molecules results in spatially varying polarizability, imparting a phase modulation to a probe laser. This process effectively creates a tunable ``molecular waveplate'' that adapts naturally to a broad spectral range. By leveraging this approach, we can generate ultrashort vortex pulses across a wide range of wavelengths.
Under optimized gas pressure and interaction length conditions, this method allows for highly efficient conversion of circularly polarized light into the desired OAM pulse, thus enabling the generation of few-cycle, high-intensity vortex beams. This molecular waveplate, which overcomes the limitations imposed by conventional optical elements, opens up new possibilities for exploring strong-field physics, ultrafast science, and other applications that require high-intensity vortex beams.
\end{abstract}                         
	
\maketitle

Optical vortices, carrying orbital angular momentum (OAM) and characterized by their helical phase fronts \cite{Allen}, have become an important topic in modern photonics. These structured beams exhibit a phase singularity at their core, where the intensity drops to zero, creating a typical doughnut-shaped profile. Their distinctive phase and intensity characteristics offer unique advantages across applications such as optical tweezers for particle trapping \cite{Paterson, MacDonald, Grier}, 
high-capacity optical communication \cite{Barreiro,Wang2012,Bozinovic}, quantum information processing \cite{Sit,Nicolas,Wang2015}, and high-resolution phase-contrast microscopy \cite{Furhapter,Tamburini,Tan}. 

The emergence of ultrashort vortex pulses has recently opened new frontiers in strong-field light-matter interactions. The combination of temporal localization and high peak power in these pulses facilitates the investigation of OAM-dependent ultrafast phenomena and nonlinear optical effects, including spin-orbit momentum conversion \cite{Fang}, helical dichroism probing \cite{Fanciulli,Rouxel}, spatiotemporal optical vortices \cite{Chong,Hancock}, and vortex algebra \cite{Hansinger}. Additionally, ultrashort vortex pulses have shown promise as driving sources for generating novel forms of structured light, such as extreme ultraviolet (EUV) vortex beams and attosecond vortex pulses \cite{HernandezGarcia,Gariepy,Gauthier,deLasHeras}. These advancements are set to improve techniques for material characterization and for the study of light-matter interactions on the attosecond timescale.

The generation of high-quality pulsed ultrashort vortex beams, however, remains a very challenging task. Traditional methods for generating (quasi-) monochromatic vortex beams involved, among others, spiral phase plates \cite{Sueda,Beijersbergen}
(SPPs) and liquid crystal spatial light modulators \cite{Forbes,Liao}
(LC-SLMs) or cylindrical lens telescopes \cite{Courtial1999}. These methods, however, suffer from inherent bandwidth limitations. When used with the broad bandwidth associated with the ultrashort pulses, the phase plates result in topological charge dispersion and azimuthal angle-dependent group delay; the spatial light modulators cause angular dispersion due to diffraction, and the cylindrical lenses introduce significant chromatic aberration. The common approach to producing broadband or few-cycle optical vortices relies on post-compression of pre-generated narrowband vortex pulses \cite{Cao2020,Xu2022,Guer2023,Feng2023}. However, maintaining the vortex phase characteristics and minimizing energy loss during compression presents a significant challenge.  

In this Letter, we present a novel approach to generating broadband optical vortices based on the non-adiabatic alignment of gas-phase molecules driven by vector beams. Molecular alignment, a well-established technique \cite{Stapelfeldt2003,Ohshima,Fleischer,Lemeshko,Koch2019,Lin2020,Xu2020,Lin2018,He2019,He2018} with applications ranging from strong-field ionization \cite{Litvinyuk2003,Thomann2008,Mikosch2013}, ultrashort pulse compression \cite{Wu2008,Bartels2001} to high harmonic generation (HHG) \cite{Itatani2004,Smirnova2009,McFarland2008,Kraus2015,He2022,He2023,He2024}, offers a powerful mechanism for spatial phase modulation. When a pulsed vector beam aligns a gas-phase molecular ensemble, it creates a spatial anisotropy in the molecular polarizability, effectively generating a `molecular waveplate'. This waveplate induces spatial phase modulation on any laser pulses passing through the aligned ensemble. We demonstrate, both theoretically and experimentally, that when a right (left) circularly polarized probe pulse interacts with such a molecular waveplate, an optical vortex beam with the opposite handedness and a topological charge of $l=\pm2p$ (where \(p\) is the polarization order number of the vector beam) is generated. The method is versatile, operating across a wide spectral range from ultraviolet to infrared, and high conversion efficiency (up to 100\%) may be achieved by optimizing gas pressure and interaction distance. Moreover, since the waveplate consists of gas-phase molecules, it will not be damaged by high laser intensity. Such a molecular waveplate paves the way for the generation of few-cycle, high-intensity vortex beams, offering new possibilities for research in strong-field physics and ultrafast science.

\begin{figure}[hbt]
	\centering
	\includegraphics[width=8.8cm]{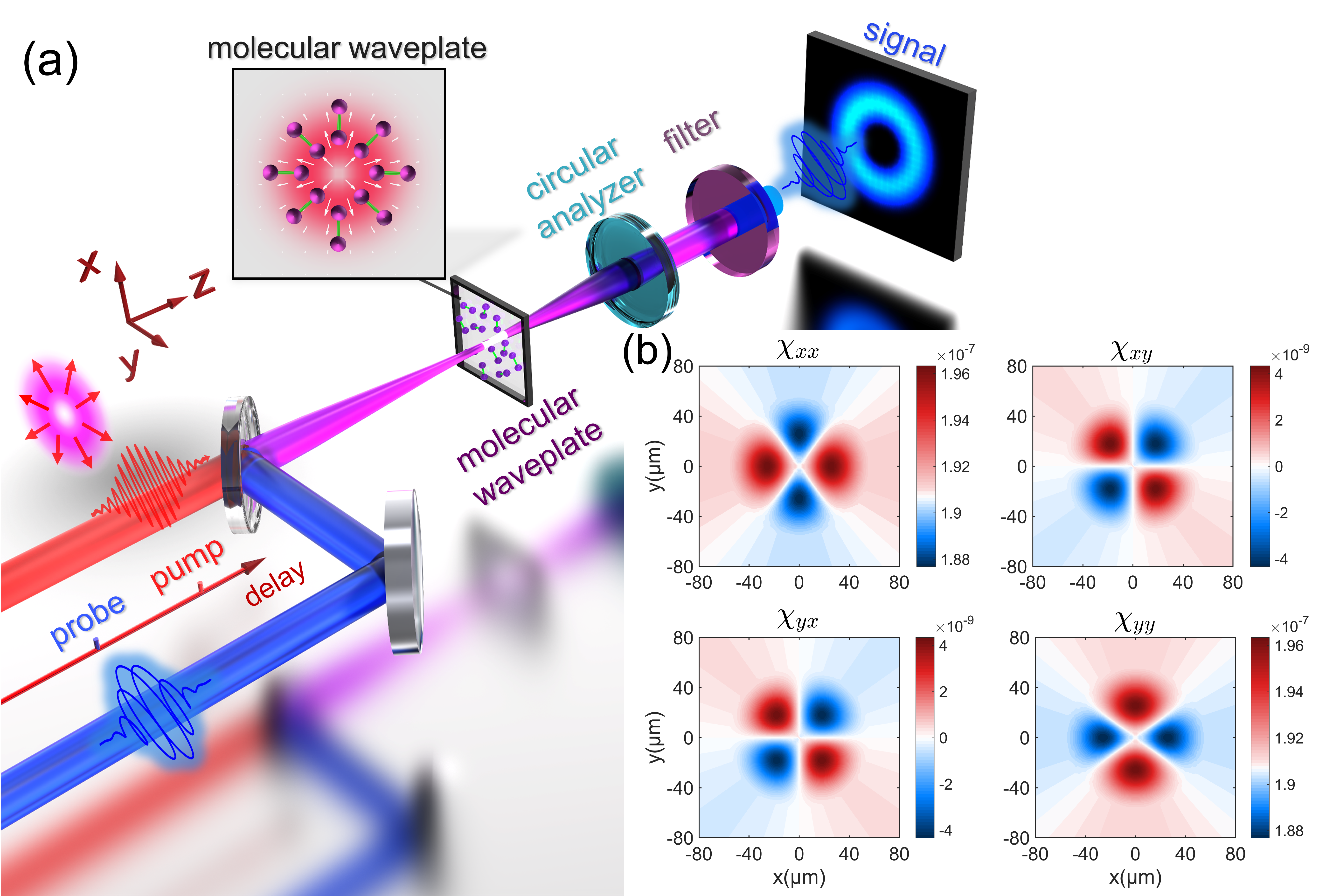}
	\caption{\label{fig1}
		(a) Schematic diagram of the `molecular waveplate'. A radially polarized light pulse induces alignment in nitrogen molecules. At specific alignment revivals, the molecular ensemble mirrors the spatial polarization pattern of the vector beam. When a circularly polarized Gaussian beam passes through this `molecular waveplate', it generates a vortex beam with the opposite circular polarization component. 
		(b) The spatial distribution of the first-order polarization tensor component is shown within the cross-section of the optical field, driven by a radially polarized pump pulse ($p=1$) with a waist radius of 50 $\mu$m and peak intensity of $6 \times 10^{13}$ W/cm$^2$ at the focal region during the N$_2$ alignment revival.} 
\end{figure}

The basic idea of our scheme is schematically shown in Fig. \ref{fig1}(a). A vector beam is used to induce non-adiabatic alignment of the molecular ensemble in time and space [Fig. 1(a)]. The Jones vector of the vector beam is given by  
$
\left[ \begin{array}{c}
	\cos \left( p\varphi +\theta _0 \right)\\
	\sin \left( p\varphi +\theta _0 \right)\\
\end{array} \right] 
$.
Here, $p$ is the polarization order number of the light field (indicating the number of polarization rotations per one optical cycle), and $\theta_0$ is the initial polarization orientation when the azimuthal angle $\varphi = 0$. At the alignment (anti-alignment) moment, the molecular ensemble is spatially aligned with the molecular axis parallel (perpendicular) to the polarization direction of the vector alignment pulse. The spatial distribution of the molecular axes gives rise to an anisotropy of molecular polarizability in space [Fig. \ref{fig1}(b)], which is similar to an optical phase waveplate. When a circularly polarized probe pulse passes through this unique `Molecular waveplate', a new optical component with opposite handedness carrying OAM is generated.

\begin{figure*}[htbp]
	\centering
	\includegraphics[width=16cm]{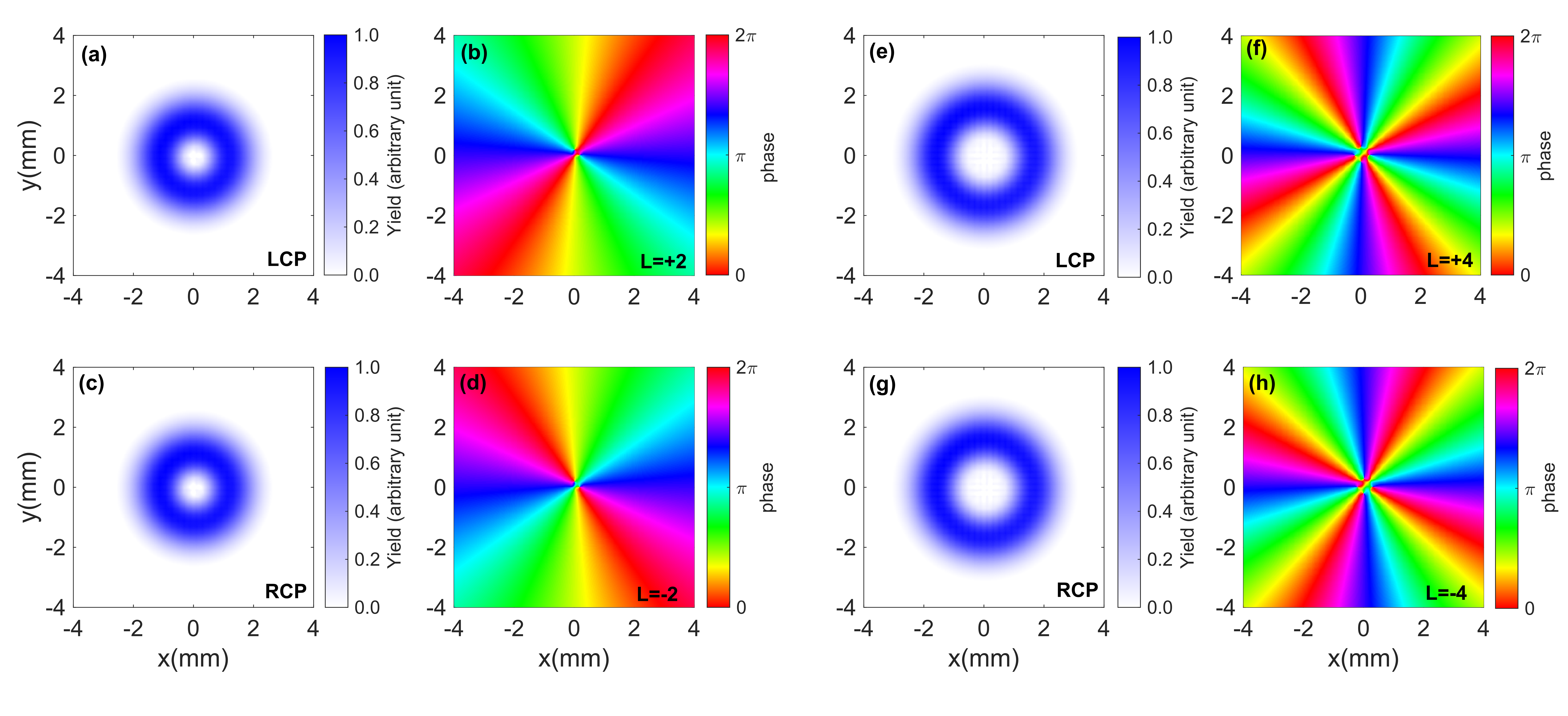}
	\caption{
		The calculated intensity profiles and spiral phases of vortex beams generated under various pump and probe pulse combinations.
		(a)-(b) A vortex beam with an OAM of  $l= +2$ is generated using a radially polarized pump pulse with a topological charge $p=1$, in conjunction with an RCP probe pulse.  (c)-(d) A vortex beam with an OAM of $l=-2$ is generated using a radially polarized pump pulse with a topological charge $p=1$ in conjunction with an LCP probe pulse. (e)-(h) Same as (a)-(d) but for a $p = 2$ radially polarized pump pulse. Here, the labels ``LCP" or ``RCP" in the intensity panels denote the handedness of the generated signal beam (The same for the figures from now on).
	}
	\label{fig2}
\end{figure*}

We first demonstrate our scheme by performing theoretical numerical simulations. The pump (alignment) pulse is a 35 fs, 800 nm, radially polarized vector beam with a polarization order number $p=1$. This vector beam is focused with a long focal length to a waist radius of 50 $\mu$m and peak intensity of $6 \times 10^{13}$ W/cm$^2$ at the focal region. The non-adiabatic molecular alignment of N$_2$ excited by this structured beam is simulated by solving the time-dependent Schr\" {o}dinger equation (TDSE) \cite{Stapelfeldt2003,Koch2019,Fleischer2009} for the molecule rotation at each spatial point in the polarization plane [i.e., the $x$-$y$ plane in Fig. \ref{fig1}(a)] of the  vector beam. Figure \ref{fig1}(b) shows the spatial distribution of the elements of the polarizability tensor simulated at the alignment moment (4.2 ps) around the half rotational revival time of N$_2$ molecules. A circularly polarized probe pulse is sent through the aligned molecules, and its interaction with the aligned molecular ensemble is simulated by solving Maxwell's propagation equations \cite{Steinitz2014} for the transverse field components ($x$ and $y$) of the probe pulse as it propagates through the prepared medium (for more details, see Supplementary Material).

Figures \ref{fig2}(a) and \ref{fig2}(b) show the simulation results for a right-handed circularly polarized (RCP) probe pulse. Figure \ref{fig2}(a) displays the intensity profile of the generated component with opposite handedness with respect to the incident probe pulse, i.e., the left-handed circularly polarized (LCP) component. The intensity profile of the newly generated LCP component exhibits a doughnut-shaped pattern with zero intensity at the center, which is a typical signature of optical-vortex beams. To assess the OAM number of the generated LCP pulse, its spatial phase is plotted in Fig. \ref{fig2}(b). As shown, the phase wraps azimuthally and anticlockwise, and it undergoes two full cycles (from 0 to 2$\pi$) over one full rotation around the optical axis. This spiral phase profile, with two cycles in one rotation, indicates the generated LCP  pulse carrying OAM of $l=+2$. Similar results can be observed for an LCP probe pulse, where an RCP pulse with an OAM number of $l=-2$ is generated [see Figs. \ref{fig2}(c) and \ref{fig2}(d)].

We have also performed simulations for a vector alignment beam with  $p=2$ [see Figs. \ref{fig2}(e)-\ref{fig2}(h)]. For an incident RCP (LCP) probe pulse, a vortex LCP (RCP) signal pulse is generated with a doubled OAM number $l=\pm4$. The sign depends on the handedness of the probe pulse.
Note that the above simulation results are calculated at the alignment moment of the molecular rotational revival. We have confirmed that the vortex beam can also be generated at the anti-alignment moment of the molecule. At the anti-alignment moment, the generated vortex field is demonstrated to have the same OAM number as at the alignment moment but with a relative $\pi$ phase rotation in space (see Fig. S2 in Supplementary Material).


Following the numerical simulations, we have conducted experiments to further validate and demonstrate the new methodology. In our experiment, the output of a Ti:sapphire laser system is divided into two parts. The stronger part is polarization-shaped to produce the vector pump pulse. The weaker part is frequency doubled and split into the probe and reference beams. The pump and probe beams are combined and focused into a gas cell filled with N$_2$ gas. After the interaction region, the pump pulse is filtered out, and the signal field is separated using a circular polarization analyzer. To visualize the OAM phases, the signal beam is further interfered at a small crossing angle with the reference beam, which has a Gaussian intensity profile. More experimental details can be found in the Supplementary Material.

\begin{figure*}[htb]
	\centering
	\includegraphics[width=16cm]{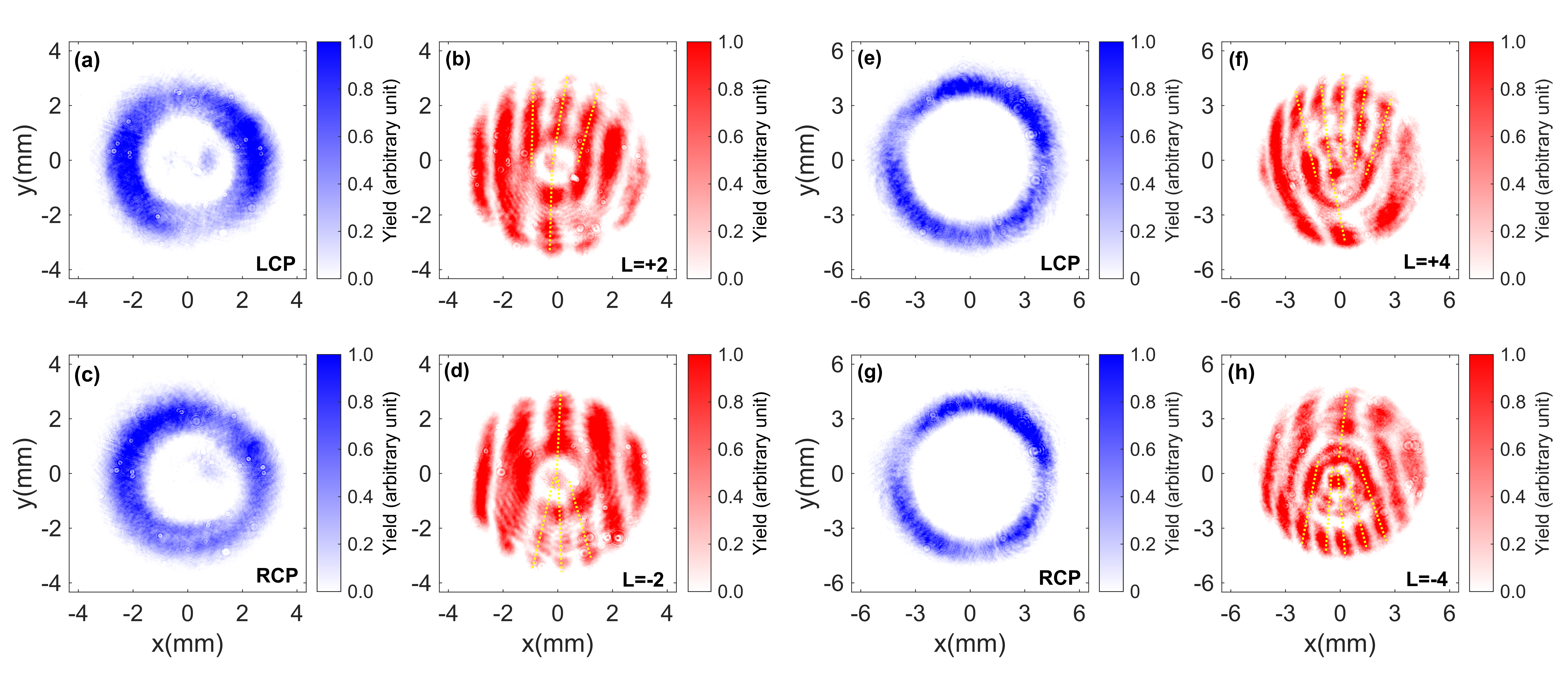}
	\caption{
		Intensity profiles of the signal field and the interferogram between the signal field and a Gaussian reference beam under various pump and probe pulse configurations.
		(a)-(b) The results when the pump pulse is a radially polarized beam with the polarization order number $p=1$, and the probe pulse is right-circularly polarized.
		(c)-(d) The results when the pump pulse is a radially polarized beam with $p=1$, and the probe pulse is left-circularly polarized. 
		(e)-(h) Same as (a)-(d) but for the $p=2$ radially polarized pump pulse. 
	}
	\label{fig3}
\end{figure*}

We first performed experiment with a $p=1$ radially polarized pump pulse. Figure \ref{fig3} depicts the results measured at the half-revival time of 4.2 ps  (alignment moment of N$_2$). Figures \ref{fig3}(a)-(d) are the intensity profile of the signal beam and its interferogram with the reference beam for RCP and LCP probe beams. As shown in Fig. \ref{fig3}(a)[\ref{fig3}(c)], a doughnut-shaped intensity profile is observed for the LCP (RCP) signal beam. Whereas Fig. \ref{fig3}(b)[\ref{fig3}(d)] shows an $l$-forked fringe pattern, which is a typical interferogram between a vortex beam and a crossing Gaussian beam. The difference in the number of fringes on the interference pattern's upper and lower sides corresponds to the optical vortex's topological charge. From Figs. \ref{fig3}(a)-\ref{fig3}(b) and Figs. \ref{fig3}(c)-\ref{fig3}(d), we can see that when an RCP (LCP) probe pulse is employed, a vortex LCP (RCP) beam with $l=\pm2$ is generated.

We have also used a pump pulse with a radially polarized vector beam with a polarization order number $p=2$ to further validate the effect of doubling the topological charge during the interaction with the aligned molecules. Figures \ref{fig3}(e)-\ref{fig3}(f) and \ref{fig3}(g)-\ref{fig3}(h) show the intensity profile of the signal pulse and its interference result with the reference beam for an RCP and LCP probe pulse, respectively. In this configuration, the generated vortex signal beams have a larger topological charge of $l=+4$ and $l=-4$, respectively. All the experimental results are in excellent agreement with the theoretical results shown in Fig. \ref{fig2}. Note that the current theoretical and experimental results are derived for a radially polarized pump pulse. In principle, our scheme can be directly applied to other vector beams. In Figs. S3-S4 in Supplementary Material, we have demonstrated the feasibility of our scheme for an azimuthally polarized pump beam.

\begin{figure}[htb]
	\centering
	\includegraphics[width=8.7cm]{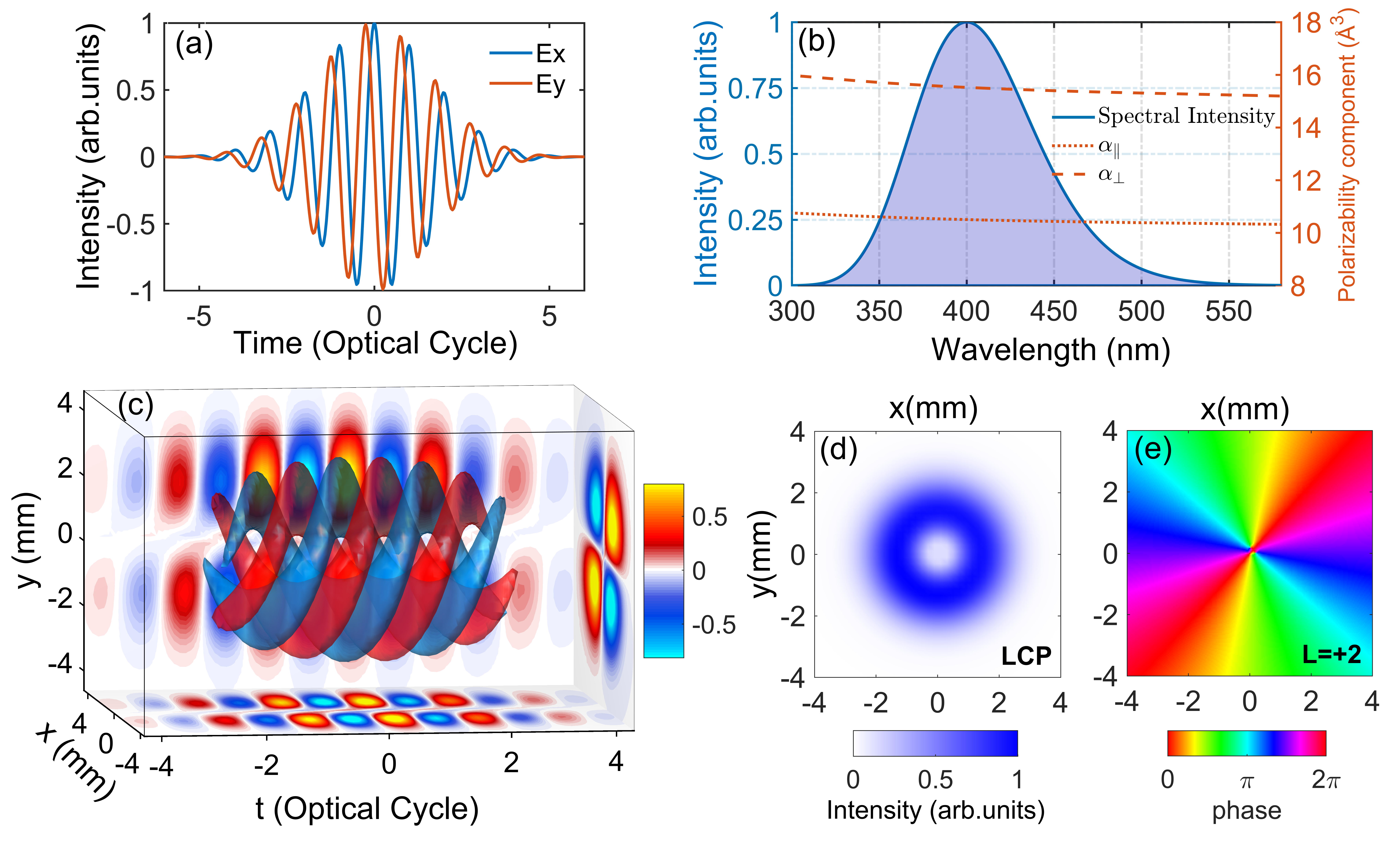}
	\caption{
		Demonstration of generating few-cycle vortex pulses using the `molecular waveplate'.
		(a) Electric field of the incident few-cycle RCP probe pulse. 
		(b) The spectrum of the few-cycle probe field and the frequency-dependent molecular polarizability components of N$_2$ molecules in this wavelength range.  
		(c) Theoretically calculated isosurfaces of the few-cycle vortex pulse electric field generated by the `molecular waveplate'. 
		(d)-(e) Intensity profiles and spiral phases at \(t=0\) of the few-cycle vortex pulse generated by the `molecular waveplate'.
	}
	\label{fig4}
\end{figure}

Unlike traditional phase modulators such as a solid spiral phase plate (SPP), the proposed molecular waveplate can well overcome the bandwidth limitations. It prevents the topological charge dispersion typically caused by the broad bandwidth of ultrashort pulses. This approach offers a significant advantage in managing the constraints imposed by the ultrashort pulses. This capability offers an opportunity for the generation of a few-cycle vortex pulse. To demonstrate this point, we have conducted numerical simulations using probe pulses with central wavelength covering a very broad spectral range: 266 nm, 400 nm and 800 nm. As shown in Fig. S5 in Supplementary Material, vortex beams can be generated with the same topological charge for these central wavelengths. In particular, we have also demonstrated the generation of a few-cycle vortex beam by our scheme. In this simulation, we used an alignment pump pulse that is a radially polarized vector beam with the polarization order number of $p=1$ and a few-cycle (three optical cycles) RCP 400 nm probe pulse. The electric field of the probe pulse, $x$ and $y$ components, are plotted in Fig. \ref{fig4}(a). Figure \ref{fig4}(b) shows the very broad spectrum of the probe field, covering the range of 300-600 nm, and the frequency-dependent molecular polarizabilities (\(\alpha_{\parallel}\) and \(\alpha_{\perp}\)) with minimal changes within the spectral range. By solving Maxwell's equations, a few-cycle LCP vortex beam with a topological charge of $l=+2$ is obtained [see Figs. \ref{fig4}(c)-\ref{fig4}(e)]. 

\begin{figure}[htb]
	\centering
	\includegraphics[width=8.5cm]{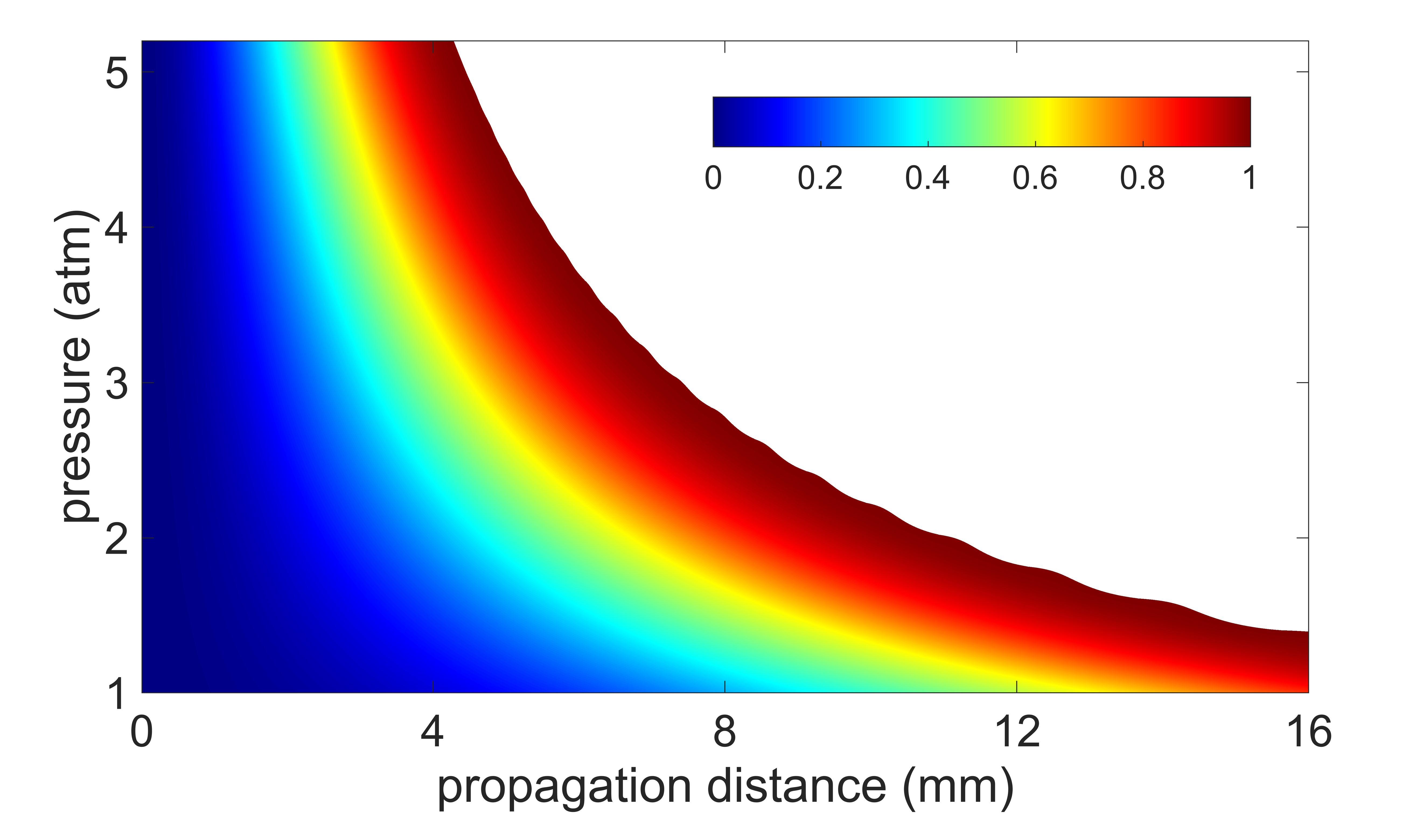}
	\caption{
		Conversion efficiency of the `molecular waveplate'.
		The color plot shows the relationship between the conversion efficiency of the generated vortex light and the gas pressure and interaction distance for a 400 nm probe pulse. The laser parameters of the pump pulse used in the simulation are the same as those in Fig. \ref{fig1}(b).
	}
	\label{fig5}
\end{figure}

Ultimately, conversion efficiency plays a crucial role in determining the feasibility of the current proposal for vortex beam generation. In our scheme, the efficiency of the vortex beam generation can be controlled by adjusting the gas pressure and the interaction length. Figure \ref{fig5} illustrates the conversion efficiency simulated for different gas pressures and interaction lengths. As shown, increasing the gas pressure and/or the interaction length can enhance conversion efficiency. Under currently available experimental conditions, vortex beam conversion efficiency approaching unity can be achieved.

In conclusion, we introduce `molecular waveplate'$-$a new approach to generating broadband, ultrashort vortex pulses. The method is based on the induced alignment of gas-phase molecules by properly polarized vector beams. This method effectively transfers the spatial polarization characteristics of the pump vector beam onto the molecular alignment distribution, leading to an anisotropic spatial polarization distribution and introducing spatial phase modulation for a probe pulse. Consequently, the prepared gas acts as a gas-phase molecular vortex plate, enabling high-efficiency vortex beam generation. The approach offers a broad operational bandwidth capable of generating vortex pulses from the ultraviolet to the mid-infrared range, producing high-quality outputs across this spectrum without encountering issues related to topological charge dispersion. Moreover, by adjusting the gas pressure and interaction distance, the vortex conversion efficiency can be approached to unity. With the tunable conversion efficiency, higher damage thresholds, and self-healing properties of gas-phase molecules, our approach demonstrates significant potential for the generation of high-power ultrashort vortex pulses, opening the way to advanced applications in nonlinear optics, including high harmonic generation and terahertz radiation.

Looking ahead, using gas-phase molecules as a programmable spatial phase modulator paves the way for applications in atmospheric OAM encoded communication, with the potential for a robust and flexible platform for data transmission. Additionally, this strategy holds promise for atmospheric vortex detection and tracking, potentially impacting the development of rotational body sensing technologies. By turning the atmosphere into a dynamic medium for structured light generation, our method sets the stage for future innovations in fundamental research and technological applications.

\begin{acknowledgements}

This work was supported by the National Key Research and Development Program of China (Grant No. 2023YFA1406800); National Natural Science Foundation of China (Grants No. 12225406, No. 12474342, No. 12074136, No. 12104349, No.12450406 and No. 12021004).
The computing work is supported by the Public Service Platform of High-Performance Computing provided by the Network and Computing Center of HUST. I.A. gratefully acknowledges the hospitality extended to him during his stay at the Department of Chemistry of the University of British Columbia. This research was made possible in part by the historic generosity of the Harold Perlman Family.
\end{acknowledgements}

			

%
	
\end{document}